\documentclass[draftcls, onecolumn]{IEEEtran}
\ifCLASSINFOpdf
\else
\fi
\usepackage{graphicx}
\usepackage{amsmath}
\usepackage{amssymb}
\usepackage{graphics}
\usepackage{latexsym}

\usepackage{tabularx}
\usepackage[nospace]{cite}

\newtheorem{Definition}{Definition}
\newtheorem{Lemma}{Lemma}
\newtheorem{Corollary}[Lemma]{Corollary}
\newtheorem{Proposition}[Lemma]{Proposition}
\newtheorem{Theorem}{Theorem}

\newtheorem{Remark}{Remark}

\usepackage[ colorlinks = true,
             linkcolor = blue,
             urlcolor  = blue,
             citecolor = red,
             anchorcolor = green,
]{hyperref}

\allowdisplaybreaks[1]

\def\Pr{{\mathrm{Pr}}}
\def\E{{\mathrm E}}
\def\Var{{\mathrm {Var}}}
\def\Cov{{\mathrm {Cov}}}

\begin{document}
%
\title{{\huge A Proof of the Strong Converse Theorem for Gaussian Broadcast Channels via the Gaussian Poincar\'{e} Inequality}}
%
%
%


\author{Silas~L.~Fong and Vincent~Y.~F.~Tan
\thanks{S.~L.~Fong and V.~Y.~F.~Tan were supported in part by National University of Singapore (NUS) Young Investigator Award under Grant R-263-000-B37-133  and in part by Ministry of Education (MOE) Tier 2 Grant R-263-000-B61-112.}%
\thanks{This paper was presented in part at the 2016 IEEE International Symposium on Information
Theory.}%
\thanks{S.~L.~Fong is with the Department of Electrical and Computer Engineering, NUS, Singapore 117583 (e-mail: \texttt{silas\_fong@nus.edu.sg}).}%
\thanks{V.~Y.~F.~Tan is with the Department of Electrical and Computer Engineering, NUS, Singapore 117583, and also with the Department of Mathematics, NUS, Singapore 119076 (e-mail: \texttt{vtan@nus.edu.sg}).}}
\maketitle
\flushbottom

\begin{abstract}
We prove that the Gaussian broadcast channel with two destinations admits the strong converse property. This implies that for every sequence of block codes operated at a common rate pair with an asymptotic average error probability $<1$, the rate pair must lie within the capacity region derived by Cover and Bergmans. The main mathematical tool required for our analysis is a logarithmic Sobolev inequality known as the Gaussian Poincar\'e inequality.
\end{abstract}

\begin{IEEEkeywords}
Gaussian broadcast channel, Gaussian Poincar\'e inequality, information spectrum, logarithmic Sobolev inequality, strong converse
\end{IEEEkeywords}

\section{Introduction}
This paper revisits the Gaussian broadcast channel (BC) \cite[Sec.~5.5]{elgamal} through which a single source would like to send information to two destinations. If the source transmits a symbol $X$, the corresponding symbols received by the two destinations are
$
Y_1 = X + Z_1$
and
$
Y_2 = X + Z_2$
respectively where $Z_1$ and $Z_2$ are zero-mean Gaussian random variables whose variances are $\sigma_1^2>0$ and $\sigma_2^2>0$ respectively. This channel is a popular model for the downlink of a cellular system. When information is sent over $n$ uses of the channel, the peak power of every transmitted codeword $(x_1, x_2, \ldots, x_n)$ is constrained to satisfy
$
\frac{1}{n}\sum_{k=1}^n x_k^2 \le P
$
for some admissible power $P>0$. Assuming that $\sigma_1^2\le\sigma_2^2$  (so the channel is degraded in favor of the first receiver), the capacity region of this channel is well known to be
\begin{equation}
\mathcal{R}_{\text{BC}}\triangleq \bigcup_{\alpha\in[0,1]}\left\{(R_1, R_2)\in \mathbb{R}_+^2\left| \parbox[c]{1.1in}{$R_1\le  \mathrm{C}\left(\frac{\alpha P}{\sigma_1^2}\right) \vspace{0.04 in }\\R_2\le \mathrm{C}\left(\frac{(1-\alpha)P}{\alpha P + \sigma_2^2}\right)$} \right.\right\}. \label{defRBC}
\end{equation}
 where
 \begin{equation}
 \mathrm{C}(x) \triangleq \frac{1}{2}\log(1+x). \label{defCx}
 \end{equation}
  The achievability part was proved by using superposition coding, an idea that originates from Cover~\cite{cover72}. The converse part was proved by Bergmans \cite{bergmans74} using the entropy power inequality~\cite{stam}. See \cite[Sec.~5.5]{elgamal} for a modern exposition of the proof of the capacity region of the Gaussian BC.


One potential drawback of the existing outer bound is the fact that it is only a {\em weak converse}, proved by using Fano's inequality~\cite[Sec.~2.1]{elgamal}. The weak converse only guarantees that the average error probabilities of any sequence of length-$n$ codes that operate at a common rate pair not belonging to the capacity region must be bounded away from~$0$ as~$n$ tends to infinity. In information theory, it is also important to establish {\em strong converse} which states that there is a sharp phase transition of asymptotic error probabilities between rate pairs inside and outside the capacity region in the following sense: Any rate pair inside the capacity region can be supported by some sequence of length-$n$ codes with asymptotic error probability being~$0$; Conversely, the asymptotic error probability of any sequence of length-$n$ codes that operate at a common rate pair not belonging to the capacity region must equal~$1$. A strong converse indicates that for any sequence of length-$n$ codes with a common rate pair that is in the exterior of the capacity region, the error probabilities must necessarily tend to~$1$.
The contrapositive of this statement can roughly be stated as follows: For any $\varepsilon \in[0,1)$ and any sequence of length-$n$ codes operated at a common rate pair that results in an asymptotic error probability not exceeding $\varepsilon$, i.e., $\varepsilon$-reliable codes, the rate pair must belong to the capacity region. This is clearly a stronger statement than the weak converse, which is a special case where $\varepsilon=0$.

\subsection{Main Contribution}
This paper provides the first formal proof of the strong converse for the Gaussian BC. We prove that for any $\varepsilon\in(0,1)$, the $\varepsilon$-capacity region of the Gaussian BC (the set containing every rate pair such that there exists a sequence of $\varepsilon$-reliable codes operated at the rate pair) is the region given in \eqref{defRBC}. In other words, if one operates at a rate pair in the exterior of the capacity region, the average error probability must necessarily tend to~$1$ as the blocklength grows. Thus, the boundary of the capacity region specifies where the sharp phase transition of asymptotic error probabilities take place.


Our technique hinges on a fundamental inequality in probability theory known as the {\em Gaussian Poincar\'e inequality}~\cite{Ledoux} (also see \cite{RagSason} and \cite{Villani}), a particular instance of a logarithmic Sobolev inequality. This inequality states that for any $n$ independent and identically distributed standard Gaussian random variables $Z^n \triangleq (Z_1,Z_2,\ldots, Z_n)$ and any differentiable mapping $f:\mathbb{R}^n\rightarrow \mathbb{R}$ where $\E[(f(Z^n))^2]<\infty$ and $\E\left[\|\nabla f(Z^n) \|^2\right]<\infty$,
\begin{equation}
\Var[f(Z^n)] \le \E\left[\|\nabla f(Z^n) \|^2\right].
\end{equation}
In Shannon theory, this inequality has been used by Polyanskiy and Verd\'u~\cite[Th.~8]{yuryOutputDistribtuion} to bound the relative entropy between the empirical distribution of an $\varepsilon$-reliable code for the additive white Gaussian noise (AWGN) channel and the $n$-fold product of the capacity-achieving output distribution. However, it has not been explicitly used in other problems in Shannon theory to establish strong converses. We find it useful in the context of the Gaussian BC to bound the variance of a certain log-likelihood ratio (information density).

An auxiliary and important contribution of our work is the following. Consider any sequence of optimal $\varepsilon$-reliable length-$n$ codes for the Gaussian BC whose rate pairs approach a specific point on the boundary of the capacity region. We show that as long as $\varepsilon<1$, those rate pairs converge to the boundary at a rate of $O\big(\frac{1}{\sqrt{n}}\big)$. The achievability part is a direct consequence of the central limit theorem, similarly to works on second-order asymptotics~\cite{TanBook} and in particular, the Gaussian multiple access channel (MAC) with degraded message sets~\cite{ScarlettTan}. However, the converse part is more involved and indeed the strong converse must first be established. The estimates obtained from the various bounding techniques contained herein, including the  Gaussian Poincar\'e inequality, allows us to assert the  $O\big(\frac{1}{\sqrt{n}}\big)$ speed of convergence. Nailing down the exact speed of convergence and the corresponding constant would be a fruitful but ambitious avenue for further research.

\subsection{Related Work}
The blowing-up lemma~\cite{Ahls76} is the standard technique for establishing the strong converses for the following network information theory problems under the discrete memoryless setting: The degraded BC~\cite[Th.~16.3]{Csi97}\cite[Th.~4]{Ahls76}, the lossless one-help-one source coding problem~\cite[Th.~16.4]{Csi97}, the MAC~\cite{dueck81}, and the Gel'fand-Pinsker channel~\cite{tyagi}. See  \cite[Sec.~3.6]{RagSason} for an exposition of the use of the blowing-up lemma for establishing the strong converse for the discrete memoryless degraded BC, and for bounding the relative entropy between the empirical distribution of good $\varepsilon$-reliable codes for the discrete memoryless channel (DMC) and the $n$-fold product of the capacity-achieving output distribution.  Similar to logarithmic Sobolev inequalities of which the Gaussian Poincar\'e inequality is a particular instance, the blowing-up lemma is a result in the study of concentration of measure~\cite{RagSason, BLM, Villani}. However, its use in Shannon theory is tailored to communication systems where the alphabets of the underlying systems are discrete (finite). It is unclear, at least to the authors, how one can adapt the use of the blowing-up lemma to establish strong converses for continuous-alphabet communication systems such as the Gaussian BC.


Another proof of the strong converse for the discrete memoryless degraded BC was recently proposed by Oohama~\cite{oohama15}. However, a crucial step in Oohama's proof relies heavily on the assumption that the input and output alphabets are finite. More specifically, in his proof of the strong converse theorem, the quantity $\xi^{''}(\lambda)$ in \cite[eq.~(20)]{oohama15} must satisfy $\lim_{\lambda \rightarrow +0}\xi^{''}(\lambda) < \infty$, which is easy to verify in the discrete memoryless case but difficult to verify in the Gaussian case.
\subsection{Paper Outline}
In the next subsection, the notation of this paper is stated. Section~\ref{sectionDefinition} contains the formulation of the Gaussian BC and our main result. Section~\ref{sectionInfoSpec} states some useful preliminary results that are used to prove the main theorem. These include some information spectrum bounds as well as an important bound based on the Gaussian Poincar\'e inequality. Section~\ref{sectionProofMainResult} presents the proof of our main result. Section~\ref{conclusion} concludes this paper. Proofs of auxiliary results are deferred to the appendices.

\subsection{Notation}\label{notation}
The sets of natural numbers, integers, real numbers and non-negative real numbers are denoted by $\mathbb{N}$, $\mathbb{Z}$, $\mathbb{R}$ and $\mathbb{R}_+$ respectively. We will take all logarithms to base~$e$ throughout this paper, so all information quantities have units of nats.
We use $\Pr\{\mathcal{E}\}$ to represent the probability of an
event~$\mathcal{E}$, and we let $\mathbf{1}\{\mathcal{E}\}$ be the indicator function of $\mathcal{E}$. A random variable is denoted by an upper-case letter (e.g., $X$), whose alphabet and realization are denoted by the corresponding calligraphic letter (e.g., $\mathcal{X}$) and lower-case letter (e.g., $x$) respectively.
We use $X^n$ to denote a random tuple $(X_1, X_2, \ldots, X_n)$, where the components $X_k$ have the same alphabet~$\mathcal{X}$.  The Euclidean norm of a tuple $x^n\in \mathbb{R}^n$ is denoted by $\|x^n\|\triangleq \sqrt{\sum_{k=1}^n x_k^2}$.

The following notations are used for any arbitrary random variables~$X$ and~$Y$ and any real-valued mapping $g$ whose domain includes $\mathcal{X}$. We let $p_X$ and $p_{Y|X}$ denote the probability distribution of $X$ and the conditional probability distribution of $Y$ given $X$ respectively.
We let $\Pr_{p_X}\{g(X)\ge\xi\}$ denote $\int_{\mathcal{X}} p_X(x)\mathbf{1}\{g(x)\ge\xi\}\, \mathrm{d}x$ for any real-valued function~$g$ and any real constant $\xi$. The expectation and the variance of~$g(X)$ are denoted as
$
\E_{p_X}[g(X)]$ and
$
 \Var_{p_X}[g(X)]\triangleq \E_{p_X}[(g(X)-\E_{p_X}[g(X)])^2]$
 respectively.
  We let $p_Xp_{Y|X}$ denote the joint distribution of $(X,Y)$, i.e., $p_Xp_{Y|X}(x,y)=p_X(x)p_{Y|X}(y|x)$ for all $x$ and $y$.
  We let $\mathcal{N}(\cdot\,;\mu,\sigma^2): \mathbb{R}^n \rightarrow \mathbb{R}_+$ be the joint probability density function of~$n$ independent copies of the Gaussian random variable whose mean and variance are~$\mu$ and~$\sigma^2$ respectively, i.e.,
   \begin{equation*}
  \mathcal{N}(z^n;\mu,\sigma^2) = \frac{1}{(2\pi \sigma^2)^{\frac{n}{2}}}e^{-\sum\limits_{k=1}^n\frac{ (z_k-\mu)^2}{2\sigma^2}}. \label{normalRVvector}
  \end{equation*}

\section{Gaussian Broadcast Channel and Its $\varepsilon$-Capacity Region} \label{sectionDefinition}
We consider the Gaussian broadcast channel (BC) where a source denoted by~$\mathrm{s}$ wants to transmit a message to two destinations denoted~$\mathrm{d}_1$ and~$\mathrm{d}_2$ respectively in~$n$ time slots (channel uses) as follows. Node~$\mathrm{s}$ chooses a message
\begin{equation}
W_i\in \{1, 2, \ldots, M_i^{(n)}\}
\end{equation}
 destined for node~$\mathrm{d}_i$ for each $i\in\{1,2\}$ where $M_i^{(n)}$ denotes the size of message $W_i$. For notational convenience, we let $\mathcal{I}\triangleq \{1,2\}$. In each time slot~$k\in\{1, 2, \ldots, n\}$, node~$\mathrm{s}$ transmits $X_k\in \mathbb{R}$ based on~$(W_1, W_2)$, and node~$\mathrm{d}_i$ receives $Y_{i,k}=X_{i,k}+Z_{i,k}$ for each $i\in\mathcal{I}$ where $\{Z_{i,k}\}_{k=1}^n$ are~$n$ independent copies of the Gaussian random variable whose mean and variance are $0$ and $\sigma_i^2$ respectively. Without loss of generality, we assume throughout the paper that
 \begin{equation}
 \sigma_2^2 \ge \sigma_1^2 >0. \label{assumptionSigma2}
 \end{equation}
  After the~$n$ time slots, node~$\mathrm{d}_i$ declares~$\hat W_i$ to be the
transmitted~$W_i$ based on $Y_i^n$ for each $i\in\mathcal{I}$. Every codeword $x^n(w_1, w_2)$ transmitted by node~$\mathrm{s}$ should always satisfy the peak power constraint $\sum_{k=1}^n x_k^2(w_1, w_2) \le n P$, where $P$ denotes the power available to node~$\mathrm{s}$. The definitions of the Gaussian BC and the codes defined on it are formally given below.
\subsection{Definitions for the Gaussian Broadcast Channel}
To simplify notation, we let $Y_\mathcal{I}\triangleq (Y_1, Y_2)$ for any random variables $(Y_1, Y_2)$, and let $\mathcal{Y}_\mathcal{I}$ and $y_\mathcal{I}$ be the alphabet and realization of $Y_\mathcal{I}$ respectively. Similarly, let $Y_{\mathcal{I},k}\triangleq (Y_{1,k}, Y_{2,k})$ for any $(Y_{1,k}, Y_{2,k})\in\mathcal{Y}_\mathcal{I}$, and let $y_{\mathcal{I},k}$ be the realization of $Y_{\mathcal{I},k}$.
\medskip
\begin{Definition} \label{defCode}
An $(n, M_\mathcal{I}^{(n)}, P)$-code, where $M_\mathcal{I}^{(n)}\triangleq (M_1^{(n)},M_2^{(n)})$, consists of the following:
\begin{enumerate}
\item A message set
$
\mathcal{W}_{i}=\{1, 2, \ldots, M_i^{(n)}\}
$
for each $i\in\mathcal{I}$.
Message $W_\mathcal{I}$ is uniform on $\mathcal{W}_\mathcal{I}$, i.e.,
\begin{equation}
\Pr\left\{W_\mathcal{I}=w_\mathcal{I}\right\}=\frac{1}{M_1^{(n)}M_2^{(n)}} \label{uniformDistribution}
\end{equation}
for all $w_\mathcal{I}\in \mathcal{W}_\mathcal{I}$ (which implies the independence between $W_1$ and $W_2$).

\item An encoding function
$
f^{(n)}: \mathcal{W}_\mathcal{I} \rightarrow \mathbb{R}^{n}
$
where $f^{(n)}$ is used by node~$\mathrm{s}$ to generate
\begin{equation}
X^n=f^{(n)}(W_\mathcal{I}). \label{encodingFunction}
 \end{equation}
 The \textit{codebook} is defined to be $\{f^{(n)}(w_\mathcal{I}) \,|\, w_\mathcal{I}\in \mathcal{W}_\mathcal{I}\}$. In addition, the peak power constraint
\begin{equation}
\|f^{(n)}(w_\mathcal{I})\|^2\le n P \label{powerConstraint}
\end{equation}
should be satisfied for each $w_\mathcal{I}\in \mathcal{W}_\mathcal{I}$.
\item A decoding function
$
\varphi_i^{(n)}: \mathbb{R}^n  \rightarrow \mathcal{W}_i
$
for each $i\in\mathcal{I}$ where $\varphi_i^{(n)}$ is used by node~$\mathrm{d}_i$ to generate
$
 \hat W_i = \varphi_i^{(n)}(Y_i^n)$.
 For each $i\in\mathcal{I}$ and each $w_i\in \mathcal{W}_i$, the \textit{decoding region of $w_i$} is defined to be
\begin{equation}
 \mathcal{D}_i^{(n)}(w_i) \triangleq \{y_i^n\in \mathbb{R}^n | \varphi_i^{(n)}(y_i^n)=w_i\}. \label{defDecodingRegion}
 \end{equation}
\end{enumerate}
\end{Definition}
\medskip
\begin{Definition}\label{defBCchannel}
The Gaussian broadcast channel (BC) is characterized by the conditional probability density function $q_{Y_\mathcal{I}|X}$ satisfying
\begin{equation}
q_{Y_\mathcal{I}|X}(y_\mathcal{I}|x) = \mathcal{N}(y_1-x; 0, \sigma_1^2)\,\mathcal{N}(y_2-x; 0, \sigma_2^2) \label{defChannelInDefinition}
\end{equation}
where $\sigma_2^2 \ge \sigma_1^2 >0$ such that the following holds for any $(n, M_\mathcal{I}^{(n)}, P)$-code: For each $k\in\{1, 2, \ldots, n\}$,
\begin{align}
p_{W_\mathcal{I}, X^n, Y_\mathcal{I}^n}
 = p_{W_\mathcal{I},X^n}\prod_{k=1}^n p_{Y_{\mathcal{I},k}|X_k}
 \label{memorylessStatement*}
\end{align}
where
\begin{equation}
p_{Y_{\mathcal{I},k}|X_k}(y_{\mathcal{I},k}|x_k) = q_{Y_\mathcal{I}|X}(y_{\mathcal{I},k}|x_k) \label{defChannelInDefinition*}
\end{equation}
for all $(x_k, y_{\mathcal{I},k})\in\mathbb{R}^3$.
\end{Definition}
\medskip

 For any $(n, M_\mathcal{I}^{(n)}, P)$-code, let $p_{W_\mathcal{I} ,X^n, Y_\mathcal{I}^n, \hat W_\mathcal{I}}$ be the joint distribution induced by the code. We can use Definition~\ref{defCode} and~\eqref{memorylessStatement*} to factorize $p_{W_\mathcal{I},X^n, Y_\mathcal{I}^n, \hat W_\mathcal{I}}$ as follows:
\begin{align}
 p_{W_\mathcal{I} ,X^n, Y_\mathcal{I}^n, \hat W_\mathcal{I}}
=p_{W_\mathcal{I},X^n}\left(\prod_{k=1}^n p_{Y_{\mathcal{I},k}|X_k}\right) p_{\hat W_1 |Y_1^n} p_{\hat W_2 |Y_2^n}.  \label{memorylessStatement}
\end{align}
\medskip
\begin{Definition} \label{defErrorProbability}
For an $(n, M_\mathcal{I}^{(n)}, P)$-code, the \textit{average probability of decoding error} is
 \begin{align}
\Pr\{\hat W_\mathcal{I} \ne W_\mathcal{I}\} = \Pr\{\cup_{i=1}^2\{Y_i^n \notin \mathcal{D}_i^{(n)}(W_i)\}\}.
\end{align}
We call an $(n, M_\mathcal{I}^{(n)}, P)$-code with average probability of decoding error no larger than $\varepsilon$ an $(n, M_\mathcal{I}^{(n)}, P, \varepsilon)_{\text{avg}}$-code. Similarly, we define \textit{maximal probability of decoding error} as
\begin{align}
 \max_{w_\mathcal{I}\in \mathcal{W}_\mathcal{I}}\Pr\{\hat W_\mathcal{I} \ne w_\mathcal{I} | W_\mathcal{I}=w_\mathcal{I}\}  = \max_{w_\mathcal{I}\in \mathcal{W}_\mathcal{I}}\Pr\{\cup_{i=1}^2\{Y_i^n \notin \mathcal{D}_i^{(n)}(w_i)\} | W_\mathcal{I}=w_\mathcal{I}\}.
\end{align}
We call an $(n, M_\mathcal{I}^{(n)}, P)$-code with maximal probability of decoding error no larger than $\varepsilon$ an $(n, M_\mathcal{I}^{(n)}, P, \varepsilon)_{\text{max}}$-code.
\end{Definition}
\medskip
\begin{Definition} \label{defAchievableRate}
Let $\varepsilon\in [0,1)$ be a real number. A rate pair $(R_1, R_2)$ is \textit{$\varepsilon$-achievable} if there exists a sequence of $(n, M_\mathcal{I}^{(n)}, P, \varepsilon_n)_{\text{avg}}$-codes such that
\begin{equation}
\liminf\limits_{n\rightarrow \infty}\frac{1}{n}\log M_i^{(n)} \ge R_i
\end{equation}
for each $i\in\mathcal{I}$ and
\begin{equation}
\limsup\limits_{n\rightarrow \infty}\varepsilon_n \le \varepsilon.
\end{equation}
\end{Definition}
\medskip
\begin{Definition}\label{defCapacity}
The $\varepsilon$-capacity region of the BC, denoted by $\mathcal{C}_\varepsilon$, is defined to be the set of $\varepsilon$-achievable rate pairs.
\end{Definition}

\subsection{Main Result}
The following theorem is the main result of this paper. The proof of this theorem  is deferred to Section~\ref{sectionProofMainResult}.
\medskip
\begin{Theorem} \label{thmMainResult}
For all $\varepsilon\in[0,1)$,
\begin{equation}
\mathcal{C}_\varepsilon  \subseteq  \mathcal{R}_{\text{BC}}
\end{equation}
where $\mathcal{R}_{\text{BC}}$ is as defined in~\eqref{defRBC}.
\end{Theorem}

It is well known \cite[Th. 5.3]{elgamal} that
\begin{equation}
\mathcal{C}_0 = \mathcal{R}_{\text{BC}}. \label{zeroCapacityForBC}
 \end{equation}
 Therefore Theorem~\ref{thmMainResult} implies the strong converse for the Gaussian BC, i.e.,  that for every $\varepsilon\in [0,1)$,
\begin{equation}
 \mathcal{C}_\varepsilon  =  \mathcal{R}_{\text{BC}}.
 \end{equation}
 Before presenting the proof of Theorem~\ref{thmMainResult}, we would like to make the following two remarks.
 \medskip
\begin{Remark} \label{remark2St}
  For each $\varepsilon\in[0,1)$ and each $\lambda \in [0,1]$, define the {\em $\lambda$-sum capacity} as
  \begin{equation}
  \mathrm{C}_\lambda \triangleq  \max_{\alpha\in[0,1]}\left\{\lambda \mathrm{C}\left(\frac{\alpha P}{\sigma_1^2}\right) + (1-\lambda)\mathrm{C}\left(\frac{(1-\alpha)P}{\alpha P + \sigma_2^2}\right)\right\}.
  \end{equation}
Theorem~\ref{thmMainResult} implies that for all $(R_1, R_2) \in\mathcal{C}_\varepsilon$,
 \begin{equation}
 \lambda R_1 + (1-\lambda) R_2 \le  \mathrm{C}_\lambda \label{remark1}
 \end{equation}
 for all $\lambda \in [0,1]$.

 In fact, our analysis gives us a useful estimate of the optimal  $\lambda$-sum rate at finite blocklengths. From the proof of Theorem~\ref{thmMainResult}, specifically the inequalities \eqref{eqn11MainProof} and \eqref{eqn12MainProof}, we may   assert the following for each $\varepsilon\in(0,1)$, each $\lambda \in [0,1]$ and each sequence of $(n, M_1^{(n)},M_2^{(n)}, P, \varepsilon)_{\text{avg}}$-codes: There exists a constant $\bar{\theta}\in\mathbb{R}$ that depends on $\varepsilon$ and $P$ (but not~$n$) such that
  \begin{equation}
\limsup_{n\to\infty}\frac{1}{\sqrt{n}} \left(\lambda \log M_1^{(n)} + (1-\lambda) \log M_2^{(n)}  - n  \mathrm{C}_\lambda \right) \le \bar{\theta}\, . \label{remark2}
 \end{equation}
 On the other hand, for each $\varepsilon\in(0,1)$, each $\lambda \in [0,1]$, it follows from the standard achievability proof involving superposition coding \cite[Ch.~5]{elgamal}  using i.i.d.\ Gaussian codewords with average power $P-\frac{1}{\sqrt{n}}$ and a generalization of   Shannon's non-asymptotic achievability bound~\cite{sha57} that there exists a sequence of $(n, M_1^{(n)},M_2^{(n)}, P, \varepsilon)_{\text{avg}}$-codes which satisfies the following: There exists a $\underline{\theta} \in\mathbb{R}$ that depends on $\varepsilon$ and $P$ (but not~$n$) such that
  \begin{equation}
\liminf_{n\to\infty}\frac{1}{\sqrt{n}} \left(\lambda \log M_1^{(n)} + (1-\lambda) \log M_2^{(n)}  - n  \mathrm{C}_\lambda\right) \ge \underline{\theta} \, .
 \end{equation}
 If we define $(M_1^*(n, \varepsilon, \lambda),M_2^*(n, \varepsilon, \lambda))$ to be an optimal pair of message sizes that satisfies
 \begin{align}
& \lambda \log M_1^*(n, \varepsilon, \lambda) + (1-\lambda) \log M_2^*(n, \varepsilon, \lambda) \notag\\
& = \max\left\{ \lambda \log M_1^{(n)} + (1-\lambda) \log M_2^{(n)}\left|\text{There exists an $(n, M_1^{(n)},M_2^{(n)}, P, \varepsilon)_{\text{avg}}$-code}\right.\right\},
 \end{align}
it then follows from \eqref{remark1} and \eqref{remark2} that
\begin{equation}
\lambda \log M_1^*(n, \varepsilon, \lambda) + (1-\lambda) \log M_2^*(n, \varepsilon, \lambda) = n \mathrm{C}_\lambda + O(\sqrt{n}) \label{eqn:lambda_sum_rate}
\end{equation}
for each $\varepsilon\in(0,1)$ and each $\lambda \in [0,1]$. This result is not unexpected in view of recent works on second-order asymptotics for network information theory problems~\cite{TanBook}. However, even establishing the strong converse is not trivial. Moreover, characterizing the order of the most significant term in the $O(\cdot)$ notation in \eqref{eqn:lambda_sum_rate} appears to be a formidable problem.
\end{Remark}
\smallskip
\begin{Remark}\label{remark1*}
As described at the beginning of this subsection, Theorem~\ref{thmMainResult} implies the strong converse under the setting of \textit{average union error probability} as defined in Definition~\ref{defErrorProbability}. Our proof technique can also be used to prove the strong converse under the setting of \textit{maximal separate error probability} as described below. Fix any $\varepsilon_1\in [0, 1)$ and any $\varepsilon_2\in [0,1)$. If we follow the setting of the discrete memoryless degraded BC in \cite[Sec. 1]{Ahls76} and define  the $(\varepsilon_1, \varepsilon_2)$-capacity region as the set of $(\varepsilon_1, \varepsilon_2)$-achievable rate pairs where
\begin{equation}
\varepsilon_i \triangleq \max_{w_\mathcal{I}\in\mathcal{W}_\mathcal{I}}\Pr\{\hat W_i \ne w_i|W_\mathcal{I}=w_\mathcal{I}\}
\end{equation}
denotes the maximal probability of decoding error for message~$i\in\mathcal{I}$, then a slight modification of the proof steps for Theorem~\ref{thmMainResult} in Section~\ref{sectionProofMainResult} (ignoring the step of codebook expurgation) will imply that the $(\varepsilon_1, \varepsilon_2)$-capacity region is contained in $\mathcal{R}_{\text{BC}}$, thus establishing the strong converse for the setting of separate maximal error probability.
\end{Remark}

\section{Preliminaries for the Proof of Theorem \ref{thmMainResult}} \label{sectionInfoSpec}
\subsection{Information Spectrum Bounds}
The following lemma is a modification of Verd\'{u}-Han's non-asymptotic converse bound \cite[Th. 4]{VH94} for obtaining a lower bound on the maximal probability of decoding error. Note that the original Verd\'{u}-Han bound pertains to the average probability of error, but the maximal probability of error is more useful in our context.
\medskip
\begin{Lemma} \label{lemmaIS}
 Fix an $(n, M_\mathcal{I}^{(n)}, P, \varepsilon)_{\text{max}}$-code with decoding regions $\{\mathcal{D}_1^{(n)}(w_1) | w_1\in\mathcal{W}_1\}$ and $\{\mathcal{D}_2^{(n)}(w_2) | w_2\in\mathcal{W}_2\}$. Let $p_{W_\mathcal{I}, X^n, Y_{\mathcal{I}}^n, \hat W_\mathcal{I}}$ denote the probability distribution induced by the code. For each $i\in\mathcal{I}$ and each $w_\mathcal{I}\in \mathcal{W}_\mathcal{I}$, fix a real number $\gamma_i(w_\mathcal{I})$. Then, we have for each $(i,j)\in \{(1,2), (2,1)\}$
 \begin{align}
 &\Pr_{p_{Y_i^n|W_\mathcal{I} = w_\mathcal{I}}}\left\{\log\frac{p_{Y_i^n|W_i}(Y_i^n|w_i)}{p_{Y_i^n}(Y_i^n)}\le \log M_i^{(n)} - \gamma_i(w_\mathcal{I})\right\} \notag\\
 &  \le  n e^{-\gamma_i(w_\mathcal{I})} + \varepsilon +  \mathbf{1}\left\{M_1^{(n)}M_2^{(n)}\!\!\int_{\mathcal{D}_i^{(n)}(w_i)}p_{Y_i^n}(y_i^n) p_{W_j|W_i, Y_i^n}(w_j|w_i, y_i^n) \mathrm{d}y_i^n > n \right\} . \label{lemmaISst2}
 \end{align}
\end{Lemma}
\begin{IEEEproof}
Fix a pair $(i,j)\in\{(1,2),(2,1)\}$, a $w_\mathcal{I}\in \mathcal{W}_\mathcal{I}$ and a real number $\gamma_i(w_\mathcal{I})$. We first consider the case where $(i,j)=(1,2)$.
In order to show \eqref{lemmaISst2} for $(i,j)=(1,2)$, we consider the following chain of inequalities where the probability terms are evaluated according to $p_{Y_1^n|W_\mathcal{I} = w_\mathcal{I}}$:
 \begin{align}
 & \Pr\left\{\log\frac{p_{Y_1^n|W_1}(Y_1^n|w_1)}{p_{Y_1^n}(Y_1^n)}\le \log M_1^{(n)} - \gamma_1(w_\mathcal{I})\right\} \notag\\
 & \le  \Pr\left\{\log\frac{p_{Y_1^n|W_1}(Y_1^n|w_1)}{p_{Y_1^n}(Y_1^n)}\le \log M_1^{(n)} - \gamma_1(w_\mathcal{I})\right\} \notag\\
 &\qquad \quad \times \mathbf{1}\left\{M_1^{(n)}M_2^{(n)}\!\!\int_{\mathcal{D}_1^{(n)}(w_1)}p_{Y_1^n}(y_1^n) p_{W_2|W_1, Y_1^n}(w_2|w_1, y_1^n) \mathrm{d}y_1^n \le n \right\} \notag\\
 &\quad +\mathbf{1}\left\{M_1^{(n)}M_2^{(n)}\!\!\int_{\mathcal{D}_1^{(n)}(w_1)}p_{Y_1^n}(y_1^n) p_{W_2|W_1, Y_1^n}(w_2|w_1, y_1^n) \mathrm{d}y_1^n > n \right\} \\
 & \le   \Pr\left\{\left\{\log\frac{p_{Y_1^n|W_1}(Y_1^n|w_1)}{p_{Y_1^n}(Y_1^n)}\le \log M_1^{(n)} - \gamma_1(w_\mathcal{I})\right\} \cap \left\{Y_1^n \in \mathcal{D}_1^{(n)}(w_1)\right\}\right\} \notag\\
  &\qquad \quad \times \mathbf{1}\left\{M_1^{(n)}M_2^{(n)}\!\!\int_{\mathcal{D}_1^{(n)}(w_1)}p_{Y_1^n}(y_1^n) p_{W_2|W_1, Y_1^n}(w_2|w_1, y_1^n) \mathrm{d}y_1^n \le n \right\} \notag\\
  & \qquad + \Pr\left\{Y_1^n \notin \mathcal{D}_1^{(n)}(w_1)\right\} +\mathbf{1}\left\{M_1^{(n)}M_2^{(n)}\!\!\int_{\mathcal{D}_1^{(n)}(w_1)}p_{Y_1^n}(y_1^n) p_{W_2|W_1, Y_1^n}(w_2|w_1, y_1^n) \mathrm{d}y_1^n > n \right\}. \label{lemmaISproof2}
  \end{align}
  In order to bound the first term in \eqref{lemmaISproof2}, we consider
  \begin{align}
  &\Pr\left\{\left\{\log\frac{p_{Y_1^n|W_1}(Y_1^n|w_1)}{p_{Y_1^n}(Y_1^n)}\le \log M_1^{(n)} - \gamma_1(w_\mathcal{I})\right\} \cap \left\{Y_1^n \in \mathcal{D}_1^{(n)}(w_1)\right\}\right\}\notag\\
  & =  \int_{\mathcal{D}_1^{(n)}(w_1)} p_{Y_1^n|W_\mathcal{I}}(y_1^n|w_\mathcal{I}) \mathbf{1}\left\{\log\frac{p_{Y_1^n|W_1}(y_1^n|w_1)}{p_{Y_1^n}(y_1^n)}\le \log M_1^{(n)} - \gamma_1(w_\mathcal{I}) \right\} \mathrm{d}y_1^n  \\
  & = M_2^{(n)}\int_{\mathcal{D}_1^{(n)}(w_1)} p_{W_2, Y_1^n|W_1}(w_2, y_1^n |w_1) \mathbf{1}\left\{\log\frac{p_{Y_1^n|W_1}(y_1^n|w_1)}{p_{Y_1^n}(y_1^n)}\le \log M_1^{(n)} - \gamma_1(w_\mathcal{I}) \right\} \mathrm{d}y_1^n \\
  & \le  M_1^{(n)}M_2^{(n)} e^{-\gamma_1(w_\mathcal{I})}\int_{\mathcal{D}_1^{(n)}(w_1)} p_{Y_1^n}(y_1^n) p_{W_2|W_1, Y_1^n}(w_2 |w_1, y_1^n) \mathrm{d}y_1^n,
  \end{align}
  which implies that
  \begin{align}
  & \Pr\left\{\left\{\log\frac{p_{Y_1^n|W_1}(Y_1^n|w_1)}{p_{Y_1^n}(Y_1^n)}\le \log M_1^{(n)} - \gamma_1(w_\mathcal{I})\right\} \cap \left\{Y_1^n \in \mathcal{D}_1^{(n)}(w_1)\right\}\right\} \notag\\
  &\qquad \quad \times \mathbf{1}\left\{M_1^{(n)}M_2^{(n)}\!\!\int_{\mathcal{D}_1^{(n)}(w_1)}p_{Y_1^n}(y_1^n) p_{W_2|W_1, Y_1^n}(w_2|w_1, y_1^n) \mathrm{d}y_1^n \le n \right\} \notag\\
 &\quad \le n e^{-\gamma_1(w_\mathcal{I})}.  \label{lemmaISproof3}
 \end{align}
The second term in \eqref{lemmaISproof2} can be upper bounded as
 \begin{align}
 \Pr\left\{Y_1^n \notin \mathcal{D}_1^{(n)}(w_1)\right\} \le \varepsilon  \label{lemmaISproof4}
 \end{align}
because the maximal probability of decoding error of the code is $\varepsilon$.
Combining \eqref{lemmaISproof2}, \eqref{lemmaISproof3} and \eqref{lemmaISproof4}, we obtain \eqref{lemmaISst2} for $(i,j)=(1,2)$. By symmetry, \eqref{lemmaISst2} holds for $(i,j)=(2,1)$.
\end{IEEEproof}
\medskip

The following corollary is a direct consequence of Lemma~\ref{lemmaIS} with an appropriate choice of $\gamma_i(w_\mathcal{I})$.
\medskip
\begin{Corollary} \label{corollaryIS}
Fix an $\varepsilon\in[0,1)$ and fix an $(n, M_\mathcal{I}^{(n)}, P, \varepsilon)_{\text{max}}$-code with decoding regions $\{\mathcal{D}_1^{(n)}(w_1) | w_1\in\mathcal{W}_1\}$ and $\{\mathcal{D}_2^{(n)}(w_2) | w_2\in\mathcal{W}_2\}$. Let $p_{W_\mathcal{I}, X^n, Y_{\mathcal{I}}^n, \hat W_\mathcal{I}}$ denote the probability distribution induced by the code. Fix a pair $(i,j)\in \{(1,2), (2,1)\}$. Then we have for each $w_\mathcal{I}\in\mathcal{W}_\mathcal{I}$
 \begin{align}
 \frac{1-\varepsilon}{2} & \le  n e^{-\left(\log M_i^{(n)}-\E\left[\log\frac{p_{Y_i^n|W_i}(Y_i^n|w_i)}{p_{Y_i^n}(Y_i^n)}\right]-\sqrt{\frac{2}{1-\varepsilon}\Var\left[\log\frac{p_{Y_i^n|W_i}(Y_i^n|w_i)}{p_{Y_i^n}(Y_i^n)}\right]}\right)} \notag\\
 &\qquad +  \mathbf{1}\left\{M_1^{(n)}M_2^{(n)}\!\!\int_{\mathcal{D}_i^{(n)}(w_i)}p_{Y_i^n}(y_i^n) p_{W_j|W_i, Y_i^n}(w_j|w_i, y_i^n) \mathrm{d}y_i^n > n \right\} \label{corollaryStatement}
 \end{align}
 where the expectation and variance terms are evaluated according to $p_{Y_i^n|W_\mathcal{I} = w_\mathcal{I}}$.
\end{Corollary}
\begin{IEEEproof}
Fix an $i\in\mathcal{I}$ and a $w_\mathcal{I}\in\mathcal{W}_\mathcal{I}$. The probability, expectation and variance terms in this proof are evaluated according to $p_{Y_i^n|W_\mathcal{I} = w_\mathcal{I}}$. Define
\begin{equation}
\gamma_i(w_\mathcal{I}) \triangleq \log M_i^{(n)}-\E\left[\log\frac{p_{Y_i^n|W_i}(Y_i^n|w_i)}{p_{Y_i^n}(Y_i^n)}\right]-\sqrt{\frac{2}{1-\varepsilon}\Var\left[\log\frac{p_{Y_i^n|W_i}(Y_i^n|w_i)}{p_{Y_i^n}(Y_i^n)}\right]}\, . \label{defGammai}
\end{equation}
Fix a pair $(i,j)\in \{(1,2), (2,1)\}$. By Chebyshev's inequality, we have for each $w_\mathcal{I}\in\mathcal{W}_\mathcal{I}$
\begin{align}
& \Pr\left\{\log\frac{p_{Y_i^n|W_i}(Y_i^n|w_i)}{p_{Y_i^n}(Y_i^n)}\le \log M_i^{(n)} - \gamma_i(w_\mathcal{I})\right\} \notag\\
 &\quad \ge 1-\frac{\Var\left[\log\frac{p_{Y_i^n|W_i}(Y_i^n|w_i)}{p_{Y_i^n}(Y_i^n)}\right]}{\left(\log M_i^{(n)} - \gamma_i(w_\mathcal{I})-\E\left[\log\frac{p_{Y_i^n|W_i}(Y_i^n|w_i)}{p_{Y_i^n}(Y_i^n)}\right]\right)^2} \, \\
 & \quad = 1-\frac{1-\varepsilon}{2} \label{corollaryInProof}
 \end{align}
 where \eqref{corollaryInProof} is due to \eqref{defGammai}.
 Combining \eqref{lemmaISst2} in Lemma~\ref{lemmaIS}, \eqref{defGammai} and \eqref{corollaryInProof}, we obtain \eqref{corollaryStatement}.
\end{IEEEproof}
\medskip

The following proposition guarantees that for any $(n, M_\mathcal{I}^{(n)}, P, \varepsilon)_{\text{max}}$-code, the last term in \eqref{corollaryStatement} equals~$1$ for only a small fraction of codewords.
\medskip
\begin{Proposition}
 \label{propositionIS*}
 Fix an $(n, M_\mathcal{I}^{(n)}, P, \varepsilon)_{\text{max}}$-code with decoding regions $\{\mathcal{D}_1^{(n)}(w_1) | w_1\in\mathcal{W}_1\}$ and $\{\mathcal{D}_2^{(n)}(w_2) | w_2\in\mathcal{W}_2\}$, and let $p_{W_\mathcal{I}, X^n, Y_{\mathcal{I}}^n, \hat W_\mathcal{I}}$ denote the probability distribution induced by the code.
For each $(i,j)\in\{(1,2), (2,1)\}$, define
\begin{align}
  {\mathcal{A}}_{(i,j)}\triangleq  \left\{w_\mathcal{I}\in \mathcal{W}_\mathcal{I}\left| M_1^{(n)}M_2^{(n)}\!\!\int_{\mathcal{D}_i^{(n)}(w_i)}p_{Y_i^n}(y_i^n) p_{W_j|W_i, Y_i^n}(w_j|w_i, y_i^n) \mathrm{d}y_i^n \le n \right.\right\}. \label{defSetA}
 \end{align}
 Then, we have for each $(i,j)\in\{(1,2), (2,1)\}$
 \begin{equation}
 \left|\mathcal{W}_\mathcal{I} \setminus {\mathcal{A}}_{(i,j)}\right| \le \frac{1}{n}\,M_1^{(n)}M_2^{(n)} \, . \label{propositionIS*st1}
 \end{equation}
 In addition, if $n\ge \frac{2}{1-\varepsilon}$, then the following holds for each $(i,j)\in\{(1,2), (2,1)\}$ and each $w_\mathcal{I}\in {\mathcal{A}}_{(i,j)}$:
 \begin{equation}
 \log M_i^{(n)} \le \E\left[\log\frac{p_{Y_i^n|W_i}(Y_i^n|w_i)}{p_{Y_i^n}(Y_i^n)}\right]+\sqrt{\frac{2}{1-\varepsilon}\Var\left[\log\frac{p_{Y_i^n|W_i}(Y_i^n|w_i)}{p_{Y_i^n}(Y_i^n)}\right]} +2\log n  \label{propositionIS*st2}
 \end{equation}
 where the expectation and variance terms are evaluated according to $p_{Y_i^n|W_\mathcal{I} = w_\mathcal{I}}$.
\end{Proposition}
\begin{IEEEproof}
Let $p_{W_\mathcal{I}, X^n, Y_{\mathcal{I}}^n, \hat W_\mathcal{I}}$ denote the probability distribution induced by the $(n, M_\mathcal{I}^{(n)}, P, \varepsilon)_{\text{max}}$-code.
Fix a pair $(i,j)\in\{(1,2), (2,1)\}$. Since $\{\mathcal{D}_i^{(n)}(w_i)| w_i\in\mathcal{W}_i\}$ consists of disjoint decoding regions by \eqref{defDecodingRegion}, we have
 \begin{align}
 &\E_{p_{W_\mathcal{I}}} \bigg[M_1^{(n)}M_2^{(n)}\!\!\int_{\mathcal{D}_i^{(n)}(W_i)}p_{Y_i^n}(y_i^n) p_{W_j|W_i, Y_i^n}(W_j|W_i, y_i^n) \mathrm{d}y_i^n \bigg]\notag\\
&\quad = \frac{1}{n}\, \sum_{w_1\in\mathcal{W}_1}\sum_{w_2\in\mathcal{W}_2}\int_{\mathcal{D}_i^{(n)}(w_i)}p_{Y_i^n}(y_i^n) p_{W_j|W_i, Y_i^n}(w_j|w_i, y_i^n) \mathrm{d}y_i^n \label{propositionISproof1(b)} \\
 &\quad=1 \label{propositionISproof0}
 \end{align}
 where the argument of the expectation is analogous to the condition in~\eqref{defSetA} that defines ${\mathcal{A}}_{(i,j)}$.
Consider the following chain of inequalities:
\begin{align}
&\Pr_{p_{W_\mathcal{I}}}\left\{W_\mathcal{I}\notin  {\mathcal{A}}_{(i,j)}\right\} \notag\\
& =  \Pr_{p_{W_\mathcal{I}}} \bigg\{M_1^{(n)}M_2^{(n)}\!\!\int_{\mathcal{D}_i^{(n)}(W_i)}p_{Y_i^n}(y_i^n) p_{W_j|W_i, Y_i^n}(W_j|W_i, y_i^n) \mathrm{d}y_i^n > n \bigg\} \\
& \le \frac{1}{n}\, \E_{p_{W_\mathcal{I}}} \bigg[M_1^{(n)}M_2^{(n)}\!\!\int_{\mathcal{D}_i^{(n)}(W_i)}p_{Y_i^n}(y_i^n) p_{W_j|W_i, Y_i^n}(W_j|W_i, y_i^n) \mathrm{d}y_i^n \bigg] \label{propositionISproof1(a)}\\
& = \frac{1}{n}
\label{propositionISproof1}
\end{align}
where
\begin{itemize}
\item \eqref{propositionISproof1(a)} is due to Markov's inequality.
\item \eqref{propositionISproof1} is due to~\eqref{propositionISproof0}.
\end{itemize}
Using \eqref{propositionISproof1} and \eqref{uniformDistribution}, we obtain \eqref{propositionIS*st1}. We will prove the second statement of the proposition in the rest of the proof. To this end, we first assume
\begin{equation}
n\ge \frac{2}{1-\varepsilon}. \label{propositionISproof2}
\end{equation}
Then for each $w_\mathcal{I}\in  {\mathcal{A}}_{(i,j)}$, it follows from Corollary~\ref{corollaryIS} and \eqref{defSetA} that
\begin{align}
\frac{1-\varepsilon}{2} & \le  n e^{-\left(\log M_i^{(n)}-\E\left[\log\frac{p_{Y_i^n|W_i}(Y_i^n|w_i)}{p_{Y_i^n}(Y_i^n)}\right]-\sqrt{\frac{2}{1-\varepsilon}\Var\left[\log\frac{p_{Y_i^n|W_i}(Y_i^n|w_i)}{p_{Y_i^n}(Y_i^n)}\right]}\right)},
\end{align}
which together with \eqref{propositionISproof2}
and implies~\eqref{propositionIS*st2}.

\end{IEEEproof}
\medskip
\subsection{The Gaussian Poincar\'{e}   Inequality}
In the proof of the main theorem, we need to use the following lemma to bound the variance term in \eqref{propositionIS*st2}, which is based on the Gaussian Poincar\'{e}  inequality. The proof of the following lemma is contained in \cite[Sec. III-C]{yuryOutputDistribtuion}, and for the sake of completeness, we provide a self-contained proof in Appendix~\ref{appendixLemmaUpperBound}.
\medskip
\begin{Lemma} \label{lemmaUpperBound}
Let $n$ be a natural number and $\sigma^2$ be a positive number.
Let $p_{W}$ be a probability distribution defined on some finite set $\mathcal{W}$, and let $g:\mathcal{W}\rightarrow \mathbb{R}^n$ be a mapping. In addition, define $p_{Z^n}$ to be the distribution of $n$ independent copies of the zero-mean Gaussian random variable with variance $\sigma^2$, i.e., $p_{Z^n}(z^n)\triangleq \mathcal{N}(z^n;0, \sigma^2)$ for all $z^n\in\mathbb{R}^n$. Suppose there exists a $\kappa\in\mathbb{R}_+$ such that
\begin{equation}
 \max_{w\in \mathcal{W}}\|g(w)\|^2 \le \kappa\,. \label{lemmaUpperBoundSt}
\end{equation}
 Then, we have
 \begin{align}
 \Var_{p_{Z^n}}\big[ \, \log \E_{p_{W}}\left[\,p_{Z^n}(Z^n + g(W))|Z^n\right]\,
\big]  \le 2\left(n+\frac{\kappa}{\sigma^2}\right). \label{lemmaUpperBoundSt1}
\end{align}
\end{Lemma}
\subsection{Simple Upper Bounds Obtained from Fano's Inequality}
Note that the upper bounds on $\log M_1^{(n)}$ and $\log M_2^{(n)}$ in Proposition~\ref{propositionIS*} do not necessarily hold for all $w_\mathcal{I}\in\mathcal{W}_\mathcal{I}$. Therefore in the proof of the main theorem, we need to obtain other upper bounds on $\log M_1^{(n)}$ and $\log M_2^{(n)}$ for those $w_\mathcal{I}\in\mathcal{W}_\mathcal{I}$ which do not satisfy the assumption in Proposition~\ref{propositionIS*}. Consequently, we need the following upper bounds on $\log M_1^{(n)}$ and $\log M_2^{(n)}$ which hold for all $w_\mathcal{I}\in\mathcal{W}_\mathcal{I}$. Since the proof of the following upper bounds is standard (by the use of Fano's inequality~\cite[Sec.~2.1]{elgamal}), it is relegated to Appendix~\ref{appendixPropositionUpperBoundFano}.
\medskip
\begin{Proposition}\label{propositionUpperBoundFano}
Fix an $(n, M_\mathcal{I}^{(n)}, P, \varepsilon)_{\text{max}}$-code.
 Then, we have for each $i\in\mathcal{I}$
 \begin{equation}
 \log M_i^{(n)} \le \frac{1}{1-\varepsilon}\left(1 + \frac{n}{2}\log\left(1+\frac{P}{\sigma_i^2}\right)\right). \label{propositionUpperBoundFanoSt}
 \end{equation}
\end{Proposition}

\section{Proof of Theorem~\ref{thmMainResult}}\label{sectionProofMainResult}
It suffices to prove $\mathcal{C}_\varepsilon \subseteq \mathcal{R}_{\text{BC}}$ for $\varepsilon \in (0,1)$ due to \eqref{zeroCapacityForBC}. Fix an arbitrary $\varepsilon\in (0,1)$ and let $(R_1, R_2)$ be an $\varepsilon$-achievable rate pair. Then there exists a sequence of $(n, M_\mathcal{I}^{(n)}, P, \varepsilon_n)_{\text{avg}}$-codes for the BC such that
\begin{equation}
\liminf\limits_{n\rightarrow \infty}\frac{1}{n}\log M_i^{(n)} \ge R_i \label{converseProofMessageSize}
\end{equation}
for each $i\in\mathcal{I}$ and
\begin{equation}
\limsup\limits_{n\rightarrow \infty}\varepsilon_n \le \varepsilon. \label{converseProofErrorPr}
\end{equation}
By \eqref{converseProofErrorPr}, there exists an $\bar \varepsilon\in (0,1)$ such that for all sufficiently large~$n$, \begin{equation}
\varepsilon_n \le \bar \varepsilon. \label{barEpsilonIntro}
\end{equation}
 By expurgating appropriate codewords from each $(n, M_\mathcal{I}^{(n)}, P, \varepsilon_n)_{\text{avg}}$-code as suggested in \cite[Problem 8.11]{elgamal}, we can obtain for each sufficiently large~$n$ an $(n, \bar M_\mathcal{I}^{(n)}, P, \sqrt{\varepsilon_n})_{\text{max}}$-code such that
\begin{equation}
 \bar M_i^{(n)} = \left\lfloor \frac{M_i^{(n)}}{n} \right\rfloor \label{converseProofMessageSize*}
\end{equation}
 for each $i\in\mathcal{I}$. Fix a sufficiently large
 \begin{align}
 n\ge \frac{2}{1-\bar \varepsilon} \label{sufficientlyLargeN}
  \end{align}
  and the corresponding $(n, \bar M_\mathcal{I}^{(n)}, P, \sqrt{\varepsilon_n})_{\text{max}}$-code such that \eqref{barEpsilonIntro} and \eqref{converseProofMessageSize*} hold.
  We will view the $(n, \bar M_\mathcal{I}^{(n)}, P, \sqrt{\varepsilon_n})_{\text{max}}$-code as an $(n, \bar M_\mathcal{I}^{(n)}, P, \sqrt{\bar \varepsilon})_{\text{max}}$-code in the rest of the proof (cf.\ \eqref{barEpsilonIntro}). Let $p_{W_\mathcal{I}, X^n, Y_{\mathcal{I}}^n, \hat W_\mathcal{I}}$ denote the probability distribution induced by the $(n, \bar M_\mathcal{I}^{(n)}, P, \sqrt{\bar \varepsilon})_{\text{max}}$-code. For each $(i,j)\in\{(1,2), (2,1)\}$, define
 \begin{align}
 \bar{\mathcal{A}}_{(i,j)}\triangleq  \left\{w_\mathcal{I}\in \mathcal{W}_\mathcal{I}\left| \bar M_1^{(n)}\bar M_2^{(n)}\!\!\int_{\mathcal{D}_i^{(n)}(w_i)}p_{Y_i^n}(y_i^n) p_{W_j|W_i, Y_i^n}(w_j|w_i, y_i^n) \mathrm{d}y_i^n \le n \right.\right\}. \label{defSetAinProof}
 \end{align}
Using Proposition~\ref{propositionIS*} and \eqref{sufficientlyLargeN}, we have for each $(i,j)\in\{(1,2), (2,1)\}$ and each $w_\mathcal{I}\in \bar{\mathcal{A}}_{(i,j)}$
 \begin{equation}
 \left|\mathcal{W}_\mathcal{I} \setminus \bar{\mathcal{A}}_{(i,j)}\right| \le \frac{1}{n}\,\bar M_1^{(n)}\bar M_2^{(n)}  \label{eqn1MainProof}
 \end{equation}
and
 \begin{equation}
 \log \bar M_i^{(n)} \le \E_{p_{Y_i^n|W_\mathcal{I} = w_\mathcal{I}}}\left[\log\frac{p_{Y_i^n|W_i}(Y_i^n|w_i)}{p_{Y_i^n}(Y_i^n)}\right]+\sqrt{\frac{2}{1-\bar \varepsilon}\Var_{p_{Y_i^n|W_\mathcal{I} = w_\mathcal{I}}}\left[\log\frac{p_{Y_i^n|W_i}(Y_i^n|w_i)}{p_{Y_i^n}(Y_i^n)}\right]} +2\log n\,.   \label{eqn2MainProof}
 \end{equation}
Following \eqref{eqn2MainProof} and letting $f^{(n)}$ be the encoding function of the $(n, \bar M_\mathcal{I}^{(n)}, P, \sqrt{\bar \varepsilon})_{\text{max}}$-code (cf.\ Definition~\ref{defCode}), we consider the following chain of inequalities for each $w_\mathcal{I}\in\mathcal{W}_\mathcal{I}$ where the variance terms are evaluated according to $p_{Y_1^n|W_\mathcal{I} = w_\mathcal{I}}$:
\begin{align}
& \Var\left[\log p_{Y_1^n|W_1}(Y_1^n|w_1)\right] \notag\\
& = \Var\left[\log \sum_{\tilde w_2\in \mathcal{W}_2}\frac{1}{\bar M_2^{(n)}} p_{Y_1^n|W_1, W_2}(Y_1^n|w_1, \tilde w_2)\right] \\
& = \Var\left[\log \sum_{\tilde w_2\in \mathcal{W}_2}\frac{1}{\bar M_2^{(n)}} p_{Y_1^n|X^n}(Y_1^n|f^{(n)}(w_1, \tilde w_2))\right] \label{eqn3MainProofStepb(a)}\\
& = \Var\left[\log \sum_{\tilde w_2\in \mathcal{W}_2}\frac{1}{\bar M_2^{(n)}} p_{Y_1^n|X^n}(Y_1^n|f^{(n)}(w_\mathcal{I})-(f^{(n)}(w_\mathcal{I})-f^{(n)}(w_1, \tilde w_2)))\right] \\
& = \int_{\mathbb{R}^n} \mathcal{N}(z^n;0, \sigma_1^2) \left( \log \sum_{\tilde w_2\in \mathcal{W}_2}\frac{1}{\bar M_2^{(n)}}
\mathcal{N}(z^n + f^{(n)}(w_\mathcal{I}) - f^{(n)}(w_1, \tilde w_2);0, \sigma_1^2)
\right)^2 \mathrm{d}z^n \nonumber\\
&\qquad - \left( \int_{\mathbb{R}^n} \mathcal{N}(z^n;0, \sigma_1^2) \log \sum_{\tilde w_2\in \mathcal{W}_2}\frac{1}{\bar M_2^{(n)}}
\mathcal{N}(z^n + f^{(n)}(w_\mathcal{I})-f^{(n)}(w_1, \tilde w_2);0, \sigma_1^2)
 \mathrm{d}z^n\right)^2 \label{eqn3MainProofStepb}
 \end{align}
 where
\begin{itemize}
\item \eqref{eqn3MainProofStepb(a)} follows from the fact due to~\eqref{encodingFunction} and~\eqref{memorylessStatement*} that for each $w_\mathcal{I}\in \mathcal{W}_\mathcal{I}$ and each $y_1^n\in\mathbb{R}^n$,
\begin{align}
p_{Y_1^n|W_\mathcal{I}}(y_1^n|w_\mathcal{I})
 = p_{Y_1^n|X^n}(y_1^n|f^{(n)}(w_\mathcal{I})). \label{pYgivenW}
\end{align}
\item \eqref{eqn3MainProofStepb} follows from letting $z^n\triangleq y_1^n - f^{(n)}(w_\mathcal{I})$ and from Definition~\ref{defBCchannel} that for each $x^n\in\mathbb{R}^n$ and each $y_1^n\in\mathbb{R}^n$,
\begin{align}
p_{Y_1^n|X^n}(y_1^n|x^n) & = \mathcal{N}(y_1^n-x^n; 0, \sigma_1^2). \label{eqnYGivenX}
\end{align}
\end{itemize}
By viewing the difference of two terms in \eqref{eqn3MainProofStepb} as
\begin{equation}
\Var_{p_{Z^n}}\left[\log \E_{p_{W_2}}\left[ p_{Z^n}(Z^n + f^{(n)}(w_\mathcal{I})-f^{(n)}(w_1, W_2))\right]\right]
\end{equation} and applying Lemma~\ref{lemmaUpperBound} (based on the Gaussian Poincar\'{e} inequality) by letting
\begin{equation}
g(\tilde w_2)\triangleq  f^{(n)}(w_\mathcal{I})-f^{(n)}(w_1, \tilde w_2)
\end{equation}
and
\begin{equation}
\kappa\triangleq \max\limits_{\tilde w_2\in\mathcal{W}_2}\|f^{(n)}(w_\mathcal{I})-f^{(n)}(w_1, \tilde w_2)\|^2,
\end{equation}
we conclude from~\eqref{eqn3MainProofStepb} that
\begin{align}
\Var\left[\log p_{Y_1^n|W_1}(Y_1^n|w_1)\right] & \le 2\left(n+ \frac{1}{\sigma_1^2}\max_{\tilde w_2\in\mathcal{W}_2}\|f^{(n)}(w_\mathcal{I})-f^{(n)}(w_1, \tilde w_2)\|^2\right) \label{eqn3MainProofStepc}\\
 & \le 2\left(n+ \frac{2}{\sigma_1^2} \left(\max_{\tilde w_2\in\mathcal{W}_2}  \|f^{(n)}(w_\mathcal{I})\|^2 + \|f^{(n)}(w_1, \tilde w_2)\|^2 \right)\right)  \label{eqn3MainProofStepd}\\
 &\le 2n\left(1+\frac{4P}{\sigma_1^2}\right),  \label{eqn3MainProof}
\end{align}
where
\begin{itemize}
    \item \eqref{eqn3MainProofStepd} follows from the fact that $\|u^n+v^n\|^2 \le 2(\|u^n\|^2 +\|v^n\|^2)$ for any $(u^n, v^n)\in \mathbb{R}^n\times \mathbb{R}^n$.
        \item \eqref{eqn3MainProof} is due to \eqref{powerConstraint}.
\end{itemize}
Following similar procedures for obtaining \eqref{eqn3MainProof}, we obtain for all $w_\mathcal{I}\in \mathcal{W}_\mathcal{I}$
\begin{align}
&\Var_{p_{Y_1^n|W_\mathcal{I} = w_\mathcal{I}}}\left[\log p_{Y_1^n}(Y_1^n)\right]  \notag\\*
&\quad=  \Var_{p_{Y_1^n|W_\mathcal{I} = w_\mathcal{I}}}\left[\log \sum_{\tilde w_\mathcal{I}\in \mathcal{W}_\mathcal{I}}\frac{1}{\bar M_1^{(n)}\bar M_2^{(n)}} p_{Y_1^n|W_\mathcal{I}}(Y_1^n|\tilde w_\mathcal{I})\right]\\
&\quad\le 2n\left(1+\frac{4P}{\sigma_1^2}\right),
\end{align}
which implies from \eqref{eqn3MainProof} that
\begin{align}
\max_{w_\mathcal{I}\in \mathcal{W}_\mathcal{I}}\max\left\{\Var_{p_{Y_1^n|W_\mathcal{I} = w_\mathcal{I}}}\left[\log p_{Y_1^n|W_1}(Y_1^n|w_1)\right], \Var_{p_{Y_1^n|W_\mathcal{I} = w_\mathcal{I}}}\left[\log p_{Y_1^n}(Y_1^n)\right]  \right\}\le 2n\left(1+\frac{4P}{\sigma_1^2}\right) . \label{eqn3MainProof*}
\end{align}
Since \eqref{eqn3MainProof*} holds, it follows by symmetry that
\begin{align}
\max_{w_\mathcal{I}\in \mathcal{W}_\mathcal{I}}\max\left\{\Var_{p_{Y_2^n|W_\mathcal{I} = w_\mathcal{I}}}\left[\log p_{Y_2^n|W_2}(Y_2^n|w_2)\right], \Var_{p_{Y_2^n|W_\mathcal{I} = w_\mathcal{I}}}\left[\log p_{Y_2^n}(Y_2^n)\right]  \right\}\le 2n\left(1+\frac{4P}{\sigma_2^2}\right) . \label{eqn3MainProof**}
\end{align}
For each $(i,j)\in\{(1,2), (2,1)\}$ and each $w_\mathcal{I}\in \bar{\mathcal{A}}_{(i,j)}$, consider
\begin{align}
& \Var_{p_{Y_i^n|W_\mathcal{I} = w_\mathcal{I}}}\left[\log\frac{p_{Y_i^n|W_i}(Y_i^n|w_i)}{p_{Y_i^n}(Y_i^n)}\right] \notag\\
& \quad \le 2 (\Var_{p_{Y_i^n|W_\mathcal{I} = w_\mathcal{I}}}\left[\log p_{Y_i^n|W_i}(Y_i^n|w_i)\right]  +\Var_{p_{Y_i^n|W_\mathcal{I} = w_\mathcal{I}}}\left[\log p_{Y_i^n}(Y_i^n)\right]) \label{eqn3MainProof**Temp(a)} \\
&\quad \le 8n\left(1+\frac{4P}{\sigma_i^2}\right) \label{eqn3MainProof**Temp(b)}
\end{align}
where
\begin{itemize}
\item \eqref{eqn3MainProof**Temp(a)} is due to the following simple fact for any real-valued random variables $U$ and $V$:
\begin{align}
\Var[U+V]
&= \Var[U] + 2 \Cov(U, V) + \Var[V] \\
&\le 2(\Var[U]+\Var[V]),  \label{eqn:cs}
\end{align}
where \eqref{eqn:cs} follows from the Cauchy-Schwartz inequality.
\item \eqref{eqn3MainProof**Temp(b)} follows from \eqref{eqn3MainProof*} and \eqref{eqn3MainProof**},
    \end{itemize}
   Combining \eqref{eqn2MainProof} and~\eqref{eqn3MainProof**Temp(b)}, we have for each $(i,j)\in\{(1,2), (2,1)\}$ and each $w_\mathcal{I}\in \bar{\mathcal{A}}_{(i,j)}$
\begin{equation}
\log \bar M_i^{(n)} \le \E_{p_{Y_i^n|W_\mathcal{I} = w_\mathcal{I}}}\left[\log\frac{p_{Y_i^n|W_i}(Y_i^n|w_i)}{p_{Y_i^n}(Y_i^n)}\right]+\sqrt{\left(\frac{16}{1-\bar \varepsilon}\right)\left(1+\frac{4P}{\sigma_i^2}\right)n} +2\log n. \label{eqn4MainProof}
\end{equation}
Consider the following chain of equalities for each $(i,j)\in\{(1,2), (2,1)\}$:
\begin{align}
&\log \bar M_i^{(n)}  \notag\\
& = \sum_{w_\mathcal{I}\in\bar{\mathcal{A}}_{(i,j)}}p_{W_\mathcal{I}}(w_\mathcal{I})\log \bar M_i^{(n)}  + \sum_{w_\mathcal{I}\in\mathcal{W}_\mathcal{I}\setminus \bar{\mathcal{A}}_{(i,j)}}p_{W_\mathcal{I}}(w_\mathcal{I})\log \bar M_i^{(n)} \\
& \le\sum_{w_\mathcal{I}\in\bar{\mathcal{A}}_{(i,j)}}p_{W_\mathcal{I}}(w_\mathcal{I})\left(\E_{p_{Y_i^n|W_\mathcal{I} = w_\mathcal{I}}}\left[\log\frac{p_{Y_i^n|W_i}(Y_i^n|w_i)}{p_{Y_i^n}(Y_i^n)}\right]+\sqrt{\left(\frac{16}{1-\bar \varepsilon}\right)\left(1+\frac{4P}{\sigma_i^2}\right)n} +2\log n\right) \notag\\
&\qquad + \sum_{w_\mathcal{I}\in\mathcal{W}_\mathcal{I}\setminus \bar{\mathcal{A}}_{(i,j)}}p_{W_\mathcal{I}}(w_\mathcal{I})\log \bar M_i^{(n)} \label{eqn4.5MainProof}\\
& \le \E_{p_{W_\mathcal{I},Y_i^n}}\left[\log\frac{p_{Y_i^n|W_i}(Y_i^n|W_i)}{p_{Y_i^n}(Y_i^n)}\right]+\sqrt{\left(\frac{16}{1-\bar \varepsilon}\right)\left(1+\frac{4P}{\sigma_i^2}\right)n} +2\log n   + \sum_{w_\mathcal{I}\in\mathcal{W}_\mathcal{I}\setminus \bar{\mathcal{A}}_{(i,j)}}p_{W_\mathcal{I}}(w_\mathcal{I}) \log \bar M_i^{(n)} \notag\\*
&\qquad - \sum_{w_\mathcal{I}\in\mathcal{W}_\mathcal{I}\setminus \bar{\mathcal{A}}_{(i,j)}}p_{W_\mathcal{I}}(w_\mathcal{I}) \left( \E_{p_{Y_i^n|W_\mathcal{I} = w_\mathcal{I}}}\left[\log\frac{p_{Y_i^n|W_i}(Y_i^n|w_i)}{p_{Y_i^n}(Y_i^n)}\right]  \right) \label{eqn4MainProof*}
\end{align}
where~\eqref{eqn4.5MainProof} is due to~\eqref{eqn4MainProof}.
In order to obtain a lower bound on $\E_{p_{Y_i^n|W_\mathcal{I} = w_\mathcal{I}}}\left[\log\frac{p_{Y_i^n|W_i}(Y_i^n|w_i)}{p_{Y_i^n}(Y_i^n)}\right]$ in \eqref{eqn4MainProof*}, we consider the following chain of inequalities for each $(i,j)\in \{(1,2), (2,1)\}$ and each $w_\mathcal{I}\in\mathcal{W}_\mathcal{I}$:
\begin{align}
&\E_{p_{Y_i^n|W_\mathcal{I} = w_\mathcal{I}}}\left[\log\frac{p_{Y_i^n|W_i}(Y_i^n|w_i)}{p_{Y_i^n}(Y_i^n)}\right] \notag\\
&\quad = \E_{p_{Y_i^n|W_\mathcal{I} = w_\mathcal{I}}}\left[\log\frac{\sum_{w_j\in\mathcal{W}_j}p_{Y_i^n, W_j|W_i}(Y_i^n, w_j|w_i)}{p_{Y_i^n}(Y_i^n)}\right] \\
&\quad \ge  \E_{p_{Y_i^n|W_\mathcal{I} = w_\mathcal{I}}}\left[\log\frac{p_{ W_j|W_i}(w_j|w_i)p_{Y_i^n|W_\mathcal{I}}(Y_i^n|w_\mathcal{I})}{p_{Y_i^n}(Y_i^n)}\right]\\
&\quad = -\log \bar M_j^{(n)} + D(p_{Y_i^n|W_\mathcal{I} = w_\mathcal{I}}\|p_{Y_i^n}) \label{eqn4MainProof****a}\\
&\quad \ge -\log \bar M_j^{(n)}. \label{eqn4MainProof****}
\end{align}
Combining \eqref{eqn4MainProof*} and \eqref{eqn4MainProof****}, we obtain for each $(i,j)\in\{(1,2), (2,1)\}$
\begin{align}
\log \bar M_i^{(n)}
& \le \E_{p_{W_\mathcal{I},Y_i^n}}\left[\log\frac{p_{Y_i^n|W_i}(Y_i^n|W_i)}{p_{Y_i^n}(Y_i^n)}\right]+\sqrt{\left(\frac{16}{1-\bar \varepsilon}\right)\left(1+\frac{4P}{\sigma_i^2}\right)n} +2\log n \notag\\*
&\qquad + \sum_{w_\mathcal{I}\in\mathcal{W}_\mathcal{I}\setminus \bar{\mathcal{A}}_{(i,j)}}p_{W_\mathcal{I}}(w_\mathcal{I})\left(\log \bar M_1^{(n)}+\log \bar M_2^{(n)}\right).\label{eqn5MainProof}
\end{align}
Following \eqref{eqn5MainProof}, we consider for each $(i,j)\in\{(1,2),(2,1)\}$
\begin{align}
& \sum_{w_\mathcal{I}\in\mathcal{W}_\mathcal{I}\setminus \bar{\mathcal{A}}_{(i,j)}}p_{W_\mathcal{I}}(w_\mathcal{I})\left(\log \bar M_1^{(n)}+\log \bar M_2^{(n)}\right) \notag\\
& \qquad \le \frac{1}{1-\bar\varepsilon}\left(\frac{2}{n } + \frac{1}{2}\log\left(1+\frac{P}{\sigma_1^2}\right)+\frac{1}{2}\log\left(1+\frac{P}{\sigma_2^2}\right)\right) \label{eqn6MainProof}
\end{align}
where the inequality follows from~\eqref{eqn1MainProof} and Proposition~\ref{propositionUpperBoundFano}.
Defining
\begin{equation}
\zeta_i\triangleq \sqrt{\left(\frac{16}{1-\bar \varepsilon}\right)\left(1+\frac{4P}{\sigma_i^2}\right)} \label{defkappai}
\end{equation}
and
\begin{equation}
\lambda_i \triangleq  2 + \frac{1}{1-\bar\varepsilon}\left(2 + \frac{1}{2}\log\left(1+\frac{P}{\sigma_1^2}\right)+\frac{1}{2}\log\left(1+\frac{P}{\sigma_2^2}\right)\right) \label{defLambdai}
\end{equation}
and identifying the fact that
\begin{equation}
\E_{p_{W_\mathcal{I},Y_i^n}}\left[\log\frac{p_{Y_i^n|W_i}(Y_i^n|W_i)}{p_{Y_i^n}(Y_i^n)}\right] = I_{p_{W_i, Y_i^n}}(W_i;Y_i^n)
\end{equation}
for each $i\in\mathcal{I}$, it follows from \eqref{eqn5MainProof} and \eqref{eqn6MainProof} that for each $i\in\mathcal{I}$
\begin{equation}
\log \bar M_i^{(n)}  \le I_{p_{W_i, Y_i^n}}(W_i;Y_i^n) + \zeta_i\sqrt{n} + \lambda_i \log n\, . \label{eqn7MainProof}
\end{equation}
For each $i\in\mathcal{I}$, since
\begin{align}
\log M_i^{(n)} \le \log (n(1+\bar M_i^{(n)})) \le \log (2n\bar M_i^{(n)}) 
\end{align}
by~\eqref{converseProofMessageSize*}, it follows from \eqref{eqn7MainProof} that
\begin{align}
\log M_i^{(n)}\le  I_{p_{W_i, Y_i^n}}(W_i;Y_i^n) + \zeta_i\sqrt{n} + (\lambda_i + 1)\log n + \log 2. \label{eqn8MainProof}
\end{align}
Following the procedures for obtaining upper bounds on $I_{p_{W_1, Y_1^n}}(W_1;Y_1^n) $ and $I_{p_{W_2, Y_2^n}}(W_2;Y_2^n) $ in the weak converse proof for the Gaussian BC \cite[Sec. 5.5.2]{elgamal}, we conclude that there exists an $\alpha_n\in[0, 1]$ such that
\begin{equation}
I_{p_{W_1, Y_1^n}}(W_1;Y_1^n) \le \frac{n}{2}\log\left(1+\frac{\alpha_n P}{\sigma_1^2}\right) \label{eqn9MainProof}
\end{equation}
and
\begin{equation}
I_{p_{W_2, Y_2^n}}(W_2;Y_2^n) \le \frac{n}{2}\log\left(1+\frac{(1-\alpha_n) P}{\alpha_n P + \sigma_2^2}\right). \label{eqn10MainProof}
\end{equation}
Combining \eqref{eqn8MainProof}, \eqref{eqn9MainProof} and \eqref{eqn10MainProof}, we obtain
\begin{equation}
\log M_1^{(n)} \le \frac{n}{2}\log\left(1+\frac{\alpha_n P}{\sigma_1^2}\right) +\zeta_1\sqrt{n} + (\lambda_1 + 1)\log n + \log 2 \label{eqn11MainProof}
\end{equation}
and
\begin{equation}
\log M_2^{(n)} \le \frac{n}{2}\log\left(1+\frac{(1-\alpha_n) P}{\alpha_n P + \sigma_2^2}\right) +\zeta_2\sqrt{n} + (\lambda_2 + 1)\log n + \log 2 \label{eqn12MainProof}
\end{equation}
where $\zeta_1$, $\zeta_2$, $\lambda_1$ and $\lambda_2$ are constants that do not depend on~$n$ by \eqref{defkappai} and \eqref{defLambdai},
which implies from \eqref{converseProofMessageSize} that
\begin{equation}
R_1 \le \liminf\limits_{n\rightarrow \infty} \frac{1}{2}\log\left(1+\frac{\alpha_n P}{\sigma_1^2}\right)
\end{equation}
and
\begin{equation}
R_2 \le \liminf\limits_{n\rightarrow \infty}\frac{1}{2}\log\left(1+\frac{(1-\alpha_n) P}{\alpha_n P + \sigma_2^2}\right),
\end{equation}
which then implies that
\begin{equation}
(R_1, R_2)\in \mathcal{R}_{\text{BC}} \label{eqn13MainProof}
\end{equation}
(cf.\ \eqref{defRBC}). Since \eqref{eqn13MainProof} holds for any $\varepsilon$-achievable $(R_1, R_2)$, it follows from Definition~\ref{defCapacity} that $\mathcal{C}_\varepsilon \subseteq \mathcal{R}_{\text{BC}}$.

\section{Concluding Remarks} \label{conclusion}
This paper provides the first formal proof of the strong converse for the Gaussian BC. Our proof technique hinges on the novel information spectrum bounds stated in Lemma~\ref{lemmaIS} and the Gaussian Poincar\'e inequality, a particular instance of a logarithmic Sobolev inequality, which leads to Lemma~\ref{lemmaUpperBound}. In addition, our proof implies that for any sequence of optimal $\varepsilon$-reliable length-$n$ codes whose rate pairs approach a specific point on the boundary of the capacity region, those rate pairs converge to the boundary at a rate of $O\big(\frac{1}{\sqrt{n}}\big)$ as long as $\varepsilon<1$ (cf.\ Remark~\ref{remark2St}). 

\appendices
\section{Proof of Lemma~\ref{lemmaUpperBound}} \label{appendixLemmaUpperBound}
Define
\begin{equation}
p_{X^n}(x^n) \triangleq \sum_{w\in\mathcal{W}}p_{W}(w)\mathbf{1}\{x^n = g(w)\} \label{lemmaUpperBoundSt*}
\end{equation}
for all $x^n \in \mathbb{R}^n$. It follows from \eqref{lemmaUpperBoundSt} and \eqref{lemmaUpperBoundSt*} that
\begin{equation}
 \max_{x^n\in \mathbb{R}^n: p_{X^n}(x^n)>0}\|x^n\|^2 \le \kappa\,. \label{lemmaUpperBoundSt**}
\end{equation}
Consider the following chain of inequalities:
\begin{align}
&\Var_{p_{Z^n}}\big[ \, \log \E_{p_{W}}\left[p_{Z^n}
(Z^n + g(W))|Z^n\right]\,
\big]  \notag\\*
&\quad =\Var_{p_{Z^n}}\big[ \, \log \E_{p_{X^n}}\left[
p_{Z^n}(Z^n + X^n)|Z^n\right]\,
\big] \label{lemmaUpperBoundEq1(a)}\\
& \quad = \int_{\mathbb{R}^n} \mathcal{N}(z^n;0, \sigma^2) \left( \log \E_{p_{X^n}}\left[
\mathcal{N}(z^n + X^n;0, \sigma^2)\right]
\right)^2 \mathrm{d}z^n \nonumber\\*
&\qquad\qquad - \left( \int_{\mathbb{R}^n} \mathcal{N}(z^n;0, \sigma^2) \log \E_{p_{X^n}}\left[
\mathcal{N}(z^n + X^n;0, \sigma^2)\right]
 \mathrm{d}z^n\right)^2 \\
 & \quad = \int_{\mathbb{R}^n} \mathcal{N}(z^n;0, 1) \left( \log \E_{p_{X^n}}\left[
\mathcal{N}\left(z^n + \frac{X^n}{\sqrt{\sigma^2}}\,;0, 1\right)\right]
\right)^2 \mathrm{d}z^n \notag \\*
&\qquad \qquad - \left( \int_{\mathbb{R}^n} \mathcal{N}(z^n;0, 1) \log \E_{p_{X^n}}\left[
\mathcal{N}\left(z^n + \frac{X^n}{\sqrt{\sigma^2}}\, ;0, 1\right)\right]
 \mathrm{d}z^n\right)^2 \\
 &\quad \le \int_{\mathbb{R}^n} \mathcal{N}(z^n;0, 1) \sum_{k=1}^n \left(\frac{\E_{p_{X^n}}\left[-\left(z_k + \frac{X_k}{\sqrt{\sigma^2}}\right)\mathcal{N}\left(z^n + \frac{X^n}{\sqrt{\sigma^2}};0, 1\right)\right]}{\E_{p_{X^n}}\left[
\mathcal{N}\left(z^n + \frac{X^n}{\sqrt{\sigma^2}};0, 1\right)\right]}\right)^2
\mathrm{d}z^n \label{lemmaUpperBoundEq1(b)} \\
& \quad =\int_{\mathbb{R}^n} \mathcal{N}(z^n;0, 1) \sum_{k=1}^n \left(-z_k -\frac{\E_{p_{X^n}}\left[\frac{X_k}{\sqrt{\sigma^2}}\mathcal{N}\left(z^n + \frac{X^n}{\sqrt{\sigma^2}};0, 1\right)\right]}{\E_{p_{X^n}}\left[
\mathcal{N}\left(z^n + \frac{X^n}{\sqrt{\sigma^2}};0, 1\right)\right]}\right)^2
\mathrm{d}z^n \\
& \quad \le 2\int_{\mathbb{R}^n} \mathcal{N}(z^n;0, 1) \sum_{k=1}^n \left( z_k^2  + \left(\frac{\E_{p_{X^n}}\left[\frac{X_k}{\sqrt{\sigma^2}}\mathcal{N}\left(z^n + \frac{X^n}{\sqrt{\sigma^2}};0, 1\right)\right]}{\E_{p_{X^n}}\left[
\mathcal{N}\left(z^n + \frac{X^n}{\sqrt{\sigma^2}};0, 1\right)\right]}\right)^2\right)
\mathrm{d}z^n \label{lemmaUpperBoundEq1(c)} \\*
&\quad = 2n + 2 \int_{\mathbb{R}^n} \mathcal{N}(z^n;0, 1) \sum_{k=1}^n \left(\frac{\E_{p_{X^n}}\left[\frac{X_k}{\sqrt{\sigma^2}}\mathcal{N}\left(z^n + \frac{X^n}{\sqrt{\sigma^2}};0, 1\right)\right]}{\E_{p_{X^n}}\left[
\mathcal{N}\left(z^n + \frac{X^n}{\sqrt{\sigma^2}};0, 1\right)\right]}\right)^2
\mathrm{d}z^n \label{lemmaUpperBoundEq1}
\end{align}
where
\begin{itemize}
\item \eqref{lemmaUpperBoundEq1(a)} follows from \eqref{lemmaUpperBoundSt*}.
\item \eqref{lemmaUpperBoundEq1(b)} follows from the Gaussian Poincar\'{e} inequality {\cite[eq.~(2.16)]{Ledoux}}, which states that for an $n$-dimensional tuple $Z^n$ consisting of independent standard Gaussian random variables and any differentiable mapping $f:\mathbb{R}^n \rightarrow \mathbb{R}$ such that $\E_{p_{Z^n}}[(f(Z^n))^2]<\infty$ and $\E_{p_{Z^n}}\left[\|\nabla f(Z^n) \|^2\right]<\infty$ where $\nabla f$ denotes the gradient of $f$,
    \begin{equation}
    \Var_{p_{Z^n}}\left[f(Z^n)\right] \le \E_{p_{Z^n}}\left[\| \nabla f(Z^n) \|^2\right].
    \end{equation}
    \item \eqref{lemmaUpperBoundEq1(c)} follows from the fact that $(a+b)^2\le 2(a^2+b^2)$ for all real numbers~$a$ and~$b$.
\end{itemize}
Following \eqref{lemmaUpperBoundEq1} and defining for each $z^n\in\mathbb{R}^n$ the distribution $\tilde p_{X^n|Z^n=z^n}$ as
\begin{equation}
\tilde p_{X^n|Z^n=z^n}(x^n) \triangleq \frac{p_{X^n}(x^n)\mathcal{N}(z^n + \frac{x^n}{\sqrt{\sigma^2}};0, 1)}{\E_{p_{X^n}}\left[\mathcal{N}\left(z^n + \frac{X^n}{\sqrt{\sigma^2}};0, 1\right)\right]}, \label{defTildePX}
\end{equation}
we consider the following chain of inequalities for each $z^n\in\mathbb{R}^n$:
\begin{align}
\sum_{k=1}^n \left(\frac{\E_{p_{X^n}}\left[\frac{X_k}{\sqrt{\sigma^2}}\mathcal{N}(z^n + \frac{X^n}{\sqrt{\sigma^2}};0, 1)\right]}{\E_{p_{X^n}}\left[
\mathcal{N}(z^n + \frac{X^n}{\sqrt{\sigma^2}} ;0, 1)\right]}\right)^2
&  = \sum_{k=1}^n \left(\E_{\tilde p_{X^n|Z^n=z^n}}\left[\frac{X_k}{\sqrt{\sigma^2}}\right]\right)^2  \\
&  \le \sum_{k=1}^n \E_{\tilde p_{X^n|Z^n=z^n}}\left[\frac{X_k^2}{\sigma^2}\right]  \\
&  \le \frac{\kappa}{\sigma^2} \label{lemmaUpperBoundEq2}
\end{align}
where the last inequality follows from \eqref{lemmaUpperBoundSt**} and \eqref{defTildePX}. Combining \eqref{lemmaUpperBoundEq1} and \eqref{lemmaUpperBoundEq2}, we obtain \eqref{lemmaUpperBoundSt1}.

\section{Proof of Proposition~\ref{propositionUpperBoundFano}}\label{appendixPropositionUpperBoundFano}
Let $p_{W_\mathcal{I}, X^n, Y_{\mathcal{I}}^n, \hat W_\mathcal{I}}$ denote the probability distribution induced by the $(n, M_\mathcal{I}^{(n)}, P, \varepsilon)_{\text{max}}$-code. For each $i\in \{1,2\}$, we follow standard procedures (for example, see~\cite[Sec.~9.2]{Cov06}) and obtain
\begin{align}
\log  M_i^{(n)} &= H(W_i)\\
& \le 1 + \varepsilon \log  M_i^{(n)}  + \sum_{k=1}^n \left(h(Y_{i,k})- h(Y_{i,k}|X_k) \right) \label{eqn6Appendix}
\end{align}
where the differential entropy terms are evaluated according to $p_{W_\mathcal{I}, X^n, Y_{\mathcal{I}}^n, \hat W_\mathcal{I}}$. On the other hand,
\begin{align}
\sum_{k=1}^n \left(h(Y_{i,k})- h(Y_{i,k}|X_k) \right) \le  \frac{n}{2}\log\left(1+\frac{P}{\sigma_i^2}\right) \label{eqn6Appendix*}
\end{align}
holds for each $i\in\mathcal{I}$ by standard arguments~\cite[Sec.~9.2]{Cov06}. Combining~\eqref{eqn6Appendix} and~\eqref{eqn6Appendix*}, we obtain
\eqref{propositionUpperBoundFanoSt}.

\section*{Acknowledgment}
The authors would like to thank Prof. Shun Watanabe for pointing out a possible extension of our result, which leads to Remark~\ref{remark1*}.



\ifCLASSOPTIONcaptionsoff
 \newpage
\fi

\end{document}